\documentclass[preprint,aps,showkeys,showpacs,floatfix]{revtex4}

\usepackage{graphicx}
\usepackage{dcolumn}
\usepackage{bm}

\usepackage{amssymb}

\newcommand{\grad}{\mbox{grad}}
\renewcommand{\div}{\mbox{div}}
\newcommand{\pdiffl}[2]{\frac{\partial #1}{\partial #2}}
\newcommand{\ldiffl}[2]{\frac{D #1}{D #2}}

\newcommand{\dfrac}[2]{\displaystyle\frac{#1}{#2}}

\begin{document}

\title{Numerical solution of shock and ramp compression \\
   for general material properties}
\author{Damian C. Swift}
\email{damian.swift@physics.org}

\affiliation{Materials Science and Technology Division, \\
   Lawrence Livermore National Laboratory, \\
   7000, East Avenue,
   Livermore, CA 94550, U.S.A.}


\begin{abstract}
A general formulation was developed to represent material models 
for applications in dynamic loading.
Numerical methods were devised to calculate response 
to shock and ramp compression, and ramp decompression,
generalizing previous solutions for scalar equations of state.
The numerical methods were found to be flexible and robust,
and matched analytic results to a high accuracy.
The basic ramp and shock solution methods were coupled to solve for
composite deformation paths, such as shock-induced impacts, and shock
interactions with a planar interface between different materials.
These calculations capture much of the physics of typical material dynamics
experiments, without requiring spatially-resolving simulations.
Example calculations were made of loading histories in metals,
illustrating the effects of plastic work on the temperatures induced
in quasi-isentropic and shock-release experiments,
and the effect of a phase transition.
\end{abstract}

\date{March 7, 2007; revised April 8, 2008 and July 1, 2008 -- LA-UR-07-2051}



\maketitle
\clearpage

\section{Introduction}
The continuum representation of matter is widely used for material dynamics
in science and engineering.
Spatially-resolved continuum dynamics simulations are the most widespread
and familiar, solving the initial value problem by discretizing the
spatial domain and integrating the dynamical equations forward in time
to predict the motion and deformation of components of the system.
This type of simulation is used, for instance, to study hypervelocity impact
problems such as the vulnerability of armor to projectiles 
\cite{Dienes70,Benson95},
the performance of satellite debris shields \cite{Gehring70}, 
and the impact of meteorites with planets, notably the formation of
the moon \cite{Canup01}.
The problem can be divided into the dynamical equations of the continuum,
the state field of the components $s(\vec r)$,
and the inherent properties of the materials.
Given the local material state $s$, the material properties allow the
stress $\tau$ to be determined.
Given the stress field $\tau(\vec r)$ and mass density field $\rho(\vec r)$,
the dynamical equations describe the fields of acceleration, compression, and
thermodynamic work done on the materials.

The equations of continuum dynamics describe the behavior of a dynamically
deforming system of arbitrary complexity.
Particular, simpler deformation paths can be described more compactly by
different sets of equations, and solved by different techniques than those 
used for continuum dynamics in general.
Simpler deformation paths occur often in experiments designed to develop and
calibrate models of material properties.
These paths can be regarded as different ways of interrogating the material
properties.
The principal examples in material dynamics are shock and ramp
compression \cite{RH,Hall01}.
Typical experiments are designed to induce such loading histories and measure
or infer the properties of the material in these states before they are 
destroyed by release from the edges or by reflected waves.

The development of the field of material dynamics was driven by applications
in the physics of hypervelocity impact and high explosive systems, including
nuclear weapons \cite{Meyers94}.
In the regimes of interest, typically components with dimensions ranging from
millimeters to meters and pressures from 1\,GPa to 1\,TPa,
material behavior is dominated by the scalar equation of state (EOS):
the relationship between pressure, compression (or mass density),
and internal energy.
Other components of stress (specifically shear stresses) are much smaller,
and chemical explosives react promptly so can be treated by simple models
of complete detonation.
EOS were developed as fits to experimental data, particularly to
series of shock states and to isothermal compression measurements
\cite{McQueen70}.
It is relatively straightforward to construct shock and ramp compression
states from an EOS algebraically or numerically depending on the EOS, 
and to fit an EOS to these measurements.
More recently, applications and scientific interest have grown 
to include a wider range of pressures and time scales,
such as laser-driven inertial confinement fusion \cite{Lindl98}, 
and experiments are designed to measure other aspects than the EOS, such
as the kinetics of phase changes, 
constitutive behavior describing shear stresses, incomplete chemical
reactions, and the effects of microstructure, including grain orientation and
porosity.
Theoretical techniques have also evolved to predict the EOS with $\sim$1\%\ 
accuracy \cite{Swift01} 
and elastic contributions to shear stress with slightly poorer
accuracy \cite{Poirier99}.

A general convention for representing material states is 
described, and numerical methods are reported for calculating shock and
ramp compression states from general representations of material properties.

\section{Conceptual structure for material properties}
The desired structure for the description of the material state and properties
under dynamic loading was developed to be as general as possible with respect to
the types of material or models to be represented in the same framework,
and designed to give the greatest amount of commonality between
spatially-resolved simulations and calculations of shock and ramp compressions.

In condensed matter on sub-microsecond time scales, heat conduction
is often too slow to have a significant effect on the response of the material,
and is ignored here.
The equations of non-relativistic continuum dynamics are,
in Lagrangian form, {\it i.e.} 
along characteristics moving with the local material velocity $\vec u(\vec r)$,
\begin{eqnarray}
\ldiffl{\rho(\vec r,t)}t & = & -\rho(\vec r,t)\div\,\vec u(\vec r,t) \\
\ldiffl{\vec u(\vec r,t)}t & = & \frac 1{\rho(\vec r,t)}\div\,\tau(\vec r,t)\\
\ldiffl{e(\vec r,t)}t & = & ||\tau(\vec r,t)\grad\,\vec u(\vec r,t)||
\end{eqnarray}
where $\rho$ is the mass density and $e$ the specific internal energy.
Changes in $e$ can be related to changes in the temperature $T$ through
the heat capacity.
The inherent properties of each material in the problem are described by its
constitutive relation or equation of state $\tau(s)$.
As well as experiencing compression and work from mechanical deformation,
the local material state $s(\vec r,t)$ can evolve through internal processes
such as plastic flow.  In general,
\begin{equation}
\ldiffl{s(\vec r,t)}t \equiv \dot s[s(\vec r,t),U(\vec r,t)]
   \quad:\quad U\equiv \grad\,\vec u(\vec r,t) \\
\end{equation}
which can also include the equations for $\partial\rho/\partial t$
and $\partial e/\partial t$.
Thus the material properties must describe at a minimum $\tau(s)$ 
and $\dot s[s(\vec r,t),U(\vec r,t)]$ for each material.
If they also describe $T(s)$, the conductivity,
and $\dot s(\dot e)$, then heat conduction can be treated.
Other functions may be needed for particular numerical methods in
continuum dynamics, such as the need for wave speeds
({\it e.g.} the longitudinal
sound speed), which are needed for time step control in explicit time
integration.
Internally, within the material properties models, it is desirable
to re-use software as much as possible, and other functions of the state
are therefore desirable to allow models to be constructed in a modular 
and hierarchical way.
Arithmetic manipulations must be performed on the state during numerical
integration,
and these can be encoded neatly using operator overloading,
so the operator of the appropriate type is invoked automatically
without having to include `if-then-else' structures for each
operator as is the case in non-object-oriented programming languages
such as Fortran-77.
For instance, if $\dot s$ is calculated in a forward-time numerical method
then changes of state are calculated using numerical evolution equations
such as
\begin{equation}
s(t+\delta t) = s(t) + \delta t \dot s.
\end{equation}
Thus for a general state $s$ and its time derivative $\dot s$,
which has an equivalent set of components, 
it is necessary to multiply a state by a real number
and to add two states together.
For a specific software implementation, other operations may be needed,
for example to create, copy, or destroy a new instance of a state.

The attraction of this approach is that, by choosing a reasonably general
form for the constitutive relation and associated operations, 
it is possible to separate the continuum dynamics part of the problem 
from the inherent behavior of the material.
The relations describing the properties of different types of material
can be encapsulated in a library form where the continuum dynamics program
need know nothing about the relations for any specific type of material,
and {\it vice versa}.
The continuum dynamics programs and the material properties relations 
can be developed and maintained independently of each other, provided that
the interface remains the same (Table~\ref{tab:interface}).
This is an efficient way to make complicated material models available
for simulations of different types, including Lagrangian and Eulerian hydrocodes
operating on different numbers of dimensions, and calculations of
specific loading or heating histories such as shock and ramp loading
discussed below.
Software interfaces have been developed in the past 
for scalar EOS with a single structure for the state \cite{AWElibs},
but object-oriented techniques make it practical to
extend the concept to much more complicated states, to combinations of models,
and to alternative types of model selected when the program is run,
without having to find a single super-set state encompassing all possible
states as special cases.

A very wide range of types of material behavior can be represented with this
formalism.
At the highest level, different types of behavior are characterized by
different structures for the state $s$ (Table~\ref{tab:stateexamples}).
For each type of state, different specific models can be defined,
such as perfect gas, polytropic and Gr\"uneisen EOS.
For each specific model, different materials are represented by choosing
different values for the parameters in the model,
and different local material states are represented through
different values for the components of $s$.
In the jargon of object-oriented programming, the ability to define an object
whose precise type is undetermined until the program is run
is known as polymorphism.
For our application, polymorphism is used at several levels in the hierarchy
of objects, from the overall type of a material
(such as `one represented by a
pressure-density-energy EOS' or `one represented by a deviatoric stress model')
through the type of relation used to describe the properties of that material
type (such as perfect gas, polytropic, or Gr\"uneisen 
for a pressure-density-energy EOS,
or Steinberg-Guinan \cite{Steinberg80} or Preston-Tonks-Wallace \cite{PTW} 
for a deviatoric stress model),
to the type of general mathematical function used to represent some
of these relations (such as a polynomial or a tabular representation
of $\gamma(\rho)$ in a polytropic EOS)
(Table~\ref{tab:polymorphism}).
States or models may be defined by extending or combining other states or models
 -- this can be implemented using the object-oriented programming concept of
inheritance.
Thus deviatoric stress models can be defined as an extension to {\it any}
pressure-density-energy EOS (they are usually written assuming a
specific type, such as Steinberg's cubic Gr\"uneisen form),
homogeneous mixtures can be defined as combinations of any
pressure-density-temperature EOS,
and heterogeneous mixtures can be defined as combinations of materials
each represented by any type of material model.

Trial implementations have been made as libraries 
in the C++ and Java programming languages \cite{software_access}.
The external interface to the material properties was general at the level
of representing a generic material type and state.
The type of state and model were then selected when programs using the
material properties library were run.
In C++, objects which were polymorphic at run time had to be represented
as pointers, requiring additional software constructions to allocate
and free up physical memory associated with each object.
It was possible to include general re-usable functions as polymorphic objects
when defining models: real functions of one real parameter could be
polynomials, transcendentals, tabular with different interpolation schemes,
piecewise definitions over different regions of the one dimensional line,
sums, products, etc; again defined specifically at run time.
Object-oriented polymorphism and inheritance were thus very powerful
techniques for increasing software re-use, making the software more
compact and more reliable through the greater use of functions which
had already been tested.

Given conceptual and software structures designed to represent general
material properties suitable for use in spatially-resolved continuum
dynamics simulations, we now consider the use of these generic material
models for calculating idealized loading paths.

\section{Idealized one-dimensional loading}
Experiments to investigate the response of materials to dynamic loading,
and to calibrate parameters in models of their behavior,
are usually designed to apply as simple a loading history as is
consistent with the transient state of interest.
The simplest canonical types of loading history are shock and ramp
\cite{RH,Hall01}.
Methods of solution are presented for calculating the result of
shock and ramp loading for materials described by generalized material
models discussed in the previous section.
Such direct solution removes the need to use a time- and space-resolved
continuum dynamics simulation, allowing states to be calculated with far
greater efficiency and without the need to consider and make allowance for
attributes of resolved simulations such as the finite numerical resolution
and the effect of numerical and artificial viscosities.

\subsection{Ramp compression}
Ramp compression is taken here to mean compression or decompression.
If the material is represented by an inviscid scalar EOS, 
{\it i.e.} 
ignoring dissipative processes and non-scalar effects from elastic strain,
ramp compression follows an isentrope.
This is no longer true when dissipative processes such as
plastic heating occur.
The term `quasi-isentropic' is sometimes used in this context,
particularly for shockless compression;
here we prefer to refer to the thermodynamic trajectories as adiabats 
since this is a more appropriate term: no heat is exchanged with
the surroundings on the time scales of interest. 

For adiabatic compression, the state evolves according to the
second law of thermodynamics,
\begin{equation}
de=T\,dS-p\,dv
\end{equation}
where $T$ is the temperature and $S$ the specific entropy.
Thus
\begin{equation}
\dot e=T\dot S-p\,\dot v = T\dot S-\frac{p\div\vec u}\rho,
\end{equation}
or for a more general material whose stress tensor is more complicated
than a scalar pressure,
\begin{equation}
de=T\,dS+\tau_n\,dv \quad\Rightarrow\quad
\dot e = T\dot S+\frac{\tau_n\div\vec u}\rho
\end{equation}
where $\tau_n$ is the component of stress normal to the direction of
deformation.
The velocity gradient was expressed through a compression factor
$\eta\equiv\rho'/\rho$ and a strain rate $\dot\epsilon$.
In all ramp experiments used in the development and calibration
of accurate material models, the strain has been applied uniaxially.
More general strain paths, for instance isotropic or 
including a shear component, can be treated by the same formalism,
and that the working rate is then a full inner product
of the stress and strain tensors.

The acceleration or deceleration of the material normal to the wave 
as it is compressed or expanded adiabatically is
\begin{equation}
\ldiffl uv = -\sqrt{-\pdiffl {\tau_n}v},
\end{equation}
from which it can be deduced that
\begin{equation}
\ldiffl u\rho = \frac{c_l}\rho
\end{equation}
where $c_l$ is the longitudinal wave speed.

As with continuum dynamics, internal evolution of the material state
can be calculated simultaneously with the continuum equations,
or operator split and calculated periodically at constant compression
\cite{Benson92}.
The results are the same to second order in the compression increment.
Operator-splitting allows calculations to be performed without an
explicit entropy, if the continuum equations are integrated isentropically
and dissipative processes are captured by internal evolution at constant
compression.

Operator-splitting is desirable when internal evolution can produce 
highly nonlinear changes, such as reaction from solid to gas:
rapid changes in state and properties can make numerical schemes 
unstable.
Operator-splitting is also desirable when the integration time step
for internal evolution is much shorter than the continuum dynamics time
step.
Neither of these considerations is very important for ramp compression
without spatial resolution, but operator-splitting was used as an option
in the ramp compression calculations for consistency with continuum
dynamics simulations.

The ramp compression equations were integrated using forward-time
Runge-Kutta numerical schemes of second order.
The fourth order scheme is a trivial extension.
The sequence of operations to calculate an increment of ramp
compression is as follows:
\begin{enumerate}
\item Time increment:
   \begin{equation}
   \delta t = -\frac{|\ln\eta|}{\dot\epsilon}
   \end{equation}
\item Predictor:
   \begin{equation}
   s(t+\delta t/2) = s(t) + \frac{\delta t}2\dot s_m(s(t),\dot\epsilon)
   \end{equation}
\item Corrector:
   \begin{equation}
   s(t+\delta t) = s(t) + \delta t\dot s_m(s(t+\delta t/2),\dot\epsilon)
   \end{equation}
\item Internal evolution:
   \begin{equation}
   s(t+\delta t) \rightarrow s(t+\delta t)
      + \int_t^{t+\delta t}\dot s_i(s(t'),\dot\epsilon)\,dt'
   \end{equation}
\end{enumerate}
where $\dot s_m$ is the model-dependent state evolution from applied strain,
and $\dot s_i$ is internal evolution at constant compression.

The independent variable for integration is specific volume $v$ or
mass density $\rho$;
for numerical integration finite steps are taken in $\rho$ and $v$.
The step size $\Delta\rho$ can be controlled so that the
numerical error during integration remains within chosen limits.
A tabular adiabat can be calculated by integrating over a range of
$v$ or $\rho$, but when simulating experimental scenarios
the upper limit for integration is usually that one of the other
thermodynamic quantities reaches a certain value, for example that
the normal component of stress reaches zero, which is the case on release from a
high pressure state at a free surface.
Specific end conditions were found by monitoring the quantity of interest
until bracketed by a finite integration step, then bisecting until
the stop condition was satisfied to a chosen accuracy.
During bisection, each trial calculation was performed as an integration 
from the first side of the bracket by the trial compression.

\subsection{Shock compression}
Shock compression is the solution of a Riemann problem for the
dynamics of a jump in compression moving with constant speed
and with a constant thickness.
The Rankine-Hugoniot (RH) equations \cite{RH} describing the shock
compression of matter are derived in the continuum approximation,
where the shock is a formal discontinuity in the continuum fields.
In reality, matter is composed of atoms, and shocks have a finite width
governed by the kinetics of dissipative processes -- at a fundamental level,
matter does not distinguish between shock compression and ramp compression
with a high strain rate -- but the RH equations apply as long
as the width of the region of matter where unresolved processes occur is 
constant.
Compared with the isentropic states induced by ramp compression in a 
material represented by an EOS, a shock always increases the entropy
and hence the temperature.
With dissipative processes included, the distinction between a ramp and a shock
may become blurred.

The RH equations express the conservation of mass, momentum, and
energy across a moving discontinuity in state.
They are usually expressed in terms of the pressure, but are readily
generalized for materials supporting shear stresses by using the 
component of stress normal to the shock ({\it i.e.}, parallel with the
direction of propagation of the shock), $\tau_n$:
\begin{eqnarray}
u_s^2 & = & -v_0^2\dfrac{\tau_n-\tau_{n0}}{v_0-v}, \label{eq:RH1} \\
\Delta u_p & = & \sqrt{-(\tau_n-\tau_{n0})(v_0-v)}, \label{eq:RH2} \\
e & = & e_0 - \frac 12 (\tau_n+\tau_{n0})(v_0-v), \label{eq:RH3}
\end{eqnarray}
where $u_s$ is the speed of the shock wave with respect to the
material, $\Delta u_p$ is the change in material
speed normal to the shock wave ({\it i.e.}, parallel to its direction
of propagation), and subscript $0$ refers to the initial state.

The RH relations can be applied to general material
models if a time scale or strain rate is imposed, and an orientation
chosen for the material with respect to the shock.
Shock compression in continuum dynamics is almost always uniaxial.

The RH equations involve only the initial and final
states in the material.
If a material has properties that depend on the deformation path
-- such as plastic flow or viscosity -- then physically the detailed shock
structure may make a difference
\cite{Ding06}.
This is a limitation of discontinuous shocks in continuum dynamics:
it may be addressed as discussed above by including dissipative
processes and considering ramp compression, if the dissipative
processes can be represented adequately in the continuum approximation.
Spatially-resolved simulations with numerical differentiation to
obtain spatial derivatives and forward time differencing are
usually not capable of representing shock discontinuities directly,
and an artificial viscosity is used to smear shock compression 
over a few spatial cells \cite{vonNeumann50}.
The trajectory followed by the material in thermodynamic space is a
smooth adiabat with dissipative heating supplied by the artificial viscosity.
If plastic work is also included during this adiabatic compression,
the overall heating for a given compression is greater than from the
RH equations.
To be consistent, plastic flow should be neglected while the
artificial viscosity is non-zero.
This localized disabling of physical processes, particularly time-dependent
ones, during the passage of the unphysically smeared shock was previously
found necessary for numerically stable simulations of detonation waves
by reactive flow \cite{rfsim}.

Detonation waves are reactive shock waves.
Steady planar detonation (the Chapman-Jouguet state \cite{Fickett}) 
may be calculated using
the RH relations, by imposing the condition that the material
state behind the shock is fully reacted.

Several numerical methods have been used to solve the RH
equations for materials represented by an EOS only
\cite{Menikoff89,Majda83}.
The general RH equations may be solved numerically for a given
shock compression $\Delta\rho$ by varying the specific internal energy $e$
until the normal stress from the material model equals that from the
RH energy equation, Eq.~\ref{eq:RH3}.
The shock and particle speeds are then calculated from Eqs~\ref{eq:RH1}
and \ref{eq:RH2}.
This numerical method is particularly convenient for EOS of the form
$p(\rho,e)$, as $e$ may be varied directly.
Solutions may still be found for general material models using
$\dot s(\dot e)$, by which the energy may be varied until the solution
is found.

Numerically, the solution was found by bracketing and bisection:
\begin{enumerate}
\item For given compression $\Delta\rho$, take the low-energy end
   for bracketing as a nearby state $s_-$ ({\it e.g.} the previous state,
   of lower compression, on the Hugoniot), compressed adiabatically
   (to state $\tilde s$),
   and cooled so the specific internal energy is $e(s_-)$.
\item Bracket the desired state: apply successively larger heating increments 
   $\Delta e$ to $\tilde s$, evolving each trial state internally,
   until $\tau_n(s)$ from the material model exceeds $\tau_n(e-e_0)$
   from Eq.~\ref{eq:RH3}.
\item Bisect in $\Delta e$, evolving each trial state internally,
   until $\tau_n(s)$ equals $\tau_n(e-e_0)$ to the desired accuracy.
\end{enumerate}

As with ramp compression, the independent variable for solution was 
mass density $\rho$,
and finite steps $\Delta\rho$ were taken.
Each shock state was calculated independently of the rest,
so numerical errors did not accumulate along the shock Hugoniot.
The accuracy of the solution was independent of $\Delta\rho$.
A tabular Hugoniot can be calculated by solving over a range of
$\rho$, but again when simulating experimental scenarios
it is usually more useful to calculate the shock state where one of the other
thermodynamic quantities reaches a certain value, often that $u_p$ and 
$\tau_n$ match the values from another, simultaneous shock calculation
for another material -- the situation in impact and shock transmission
problems, discussed below.
Specific stop conditions were found by monitoring the quantity of interest
until bracketed by a finite solution step, then bisecting until
the stop condition was satisfied to a chosen accuracy.
During bisection, each trial calculation was performed as a shock from
the initial conditions to the trial shock compression.

\subsection{Accuracy: application to air}
\begin{figure}
\begin{center}
\includegraphics[width=\textwidth]{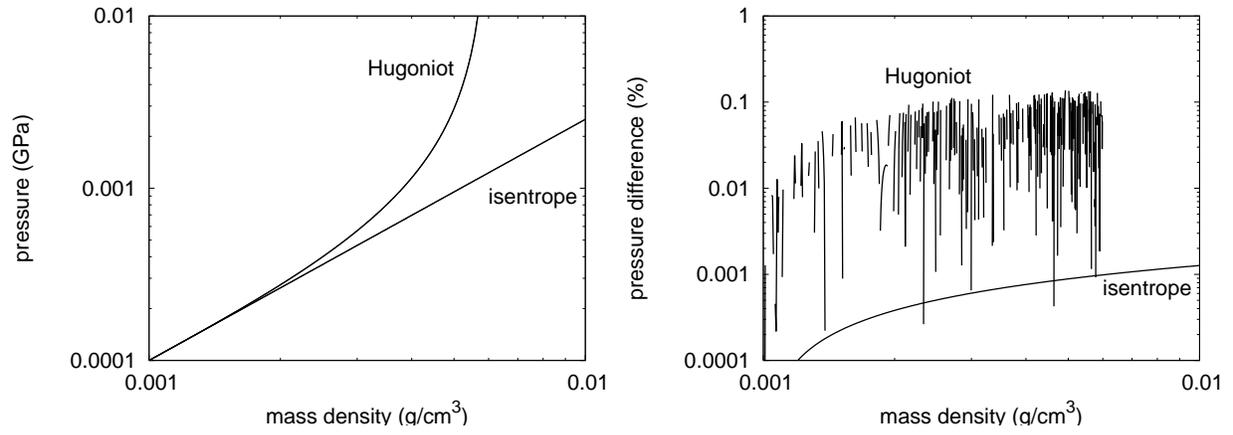}
\end{center}
\caption{Principal isentrope and shock Hugoniot for air (perfect gas):
   numerical calculations for general material models,
   compared with analytic solutions.}
\label{fig:pgascmp}
\end{figure}

The accuracy of these numerical schemes was tested by comparing with
shock and ramp compression of a material represented by a perfect gas EOS,
\begin{equation}
p=(\gamma-1)\rho e.
\end{equation}
The numerical solution requires a value to be chosen for every parameter
in the material model, here $\gamma$.
Air was chosen as an example material, with $\gamma=1.4$.
Air at standard temperature and pressure has approximately
$\rho=10^{-3}$\,g/cm$^3$ and $e=0.25$\,MJ/kg.
Isentropes for the perfect gas EOS have the form
\begin{equation}
p\rho^{-\gamma} =\quad\mbox{constant},
\end{equation}
and shock Hugoniots have the form
\begin{equation}
p = (\gamma-1)
\frac{2e_0\rho_0\rho + p_0(\rho-\rho_0)}{(\gamma+1)\rho_0-(\gamma-1)\rho}.
\end{equation}
The numerical solutions reproduced the principal isentrope and Hugoniot
to $10^{-3}$\%\ and 0.1\%\ respectively,
for a compression increment of 1\%\ along the isentrope
and a solution tolerance of $10^{-6}$\,GPa for each shock state 
(Fig.~\ref{fig:pgascmp}).
Over most of the range, the error in the Hugoniot was 0.02\%\ or less,
only approaching 0.1\%\ near the maximum shock compression.

\section{Complex behavior of condensed matter}
The ability to calculate shock and ramp loci in state space,
{\it i.e.} 
as a function of varying loading conditions, is particularly convenient
for investigating complex aspects of the response of condensed matter
to dynamic loading.
Each locus can be obtained by a single series of shock or ramp solutions,
rather than having to perform a series of time- and space-resolved 
continuum dynamics simulations, varying the initial or boundary conditions
and reducing the solution.
We consider the calculation of temperature in the scalar EOS,
the effect of material strength and the effect of phase changes.

\subsection{Temperature}
The continuum dynamics equations can be closed using a mechanical EOS
relating stress to mass density, strain, and internal energy.
For a scalar EOS, the ideal form to close the continuum equations is
$p(\rho,e)$, with $s=\{\rho,e\}$ the natural choice for the primitive state
fields.
However, the temperature is needed as a parameter in physical descriptions
of many contributions to the constitutive response, including
plastic flow, phase transitions, and chemical reactions.
Here, we discuss the calculation of temperature in different forms of the 
scalar EOS.

\subsubsection{Density-temperature equations of state}
If the scalar EOS is constructed from its underlying physical contributions
for continuum dynamics,
it may take the form $e(\rho,T)$, from which $p(\rho,T)$ can be 
calculated using the second law of thermodynamics \cite{Swift01}.
An example is the `SESAME' form of EOS, based on interpolated 
tabular relations for $\{p,e\}(\rho,T)$ \cite{SESAME}.
A pair of relations $\{p,e\}(\rho,T)$ can be used as a mechanical EOS
by eliminating $T$, which is equivalent to inverting $e(\rho,T)$ to find
$T(\rho,e)$, then substituting in $p(\rho,T)$.
For a general $e(\rho,T)$ relation, for example for the SESAME EOS,
the inverse can be calculated numerically as required, along an isochore.
In this way, a $\{p,e\}(\rho,T)$ can be used as a $p(\rho,e)$ EOS.

Alternatively, the same $p(\rho,T)$ relation can be used directly
with a primitive state field including temperature instead of energy: 
$s=\{\rho,T\}$.
The evolution of the state under mechanical work then involves the
calculation of $\dot T(\dot e)$, {\it i.e.} the reciprocal of the
specific heat capacity, which is a derivative of $e(\rho,T)$.
As this calculation does not require $e(\rho,T)$ to be inverted,
it is computationally more efficient to use  $\{p,e\}(\rho,T)$ EOS
with a temperature-based, rather than energy-based, state.
The main disadvantage is that it is more difficult to ensure exact
energy conservation as the continuum dynamics equations are integrated in time,
but any departure from exact conservation is at the level of accuracy of
the algorithm used to integrate the heat capacity.

\begin{figure}
\begin{center}\includegraphics[scale=0.80]{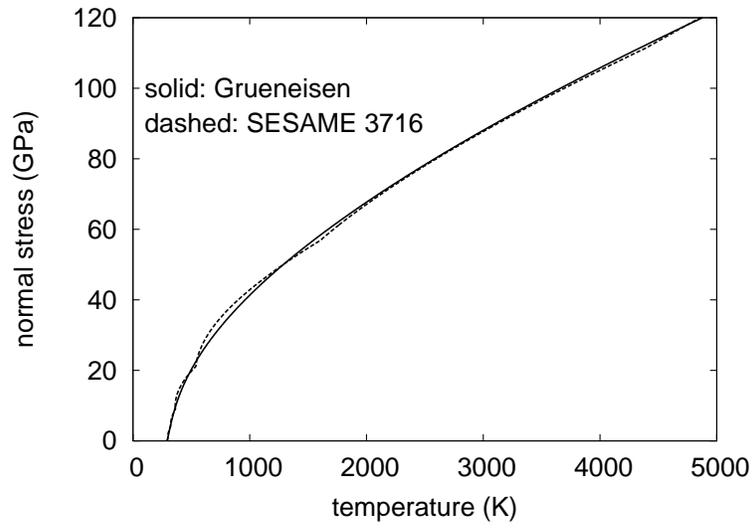}\end{center}
\caption{Shock Hugoniot for Al in pressure-temperature space,
   for different representations of the equation of state.}
\label{fig:AlcmpT}
\end{figure}

Both structures of EOS have been implemented for material property calculations.
Taking a SESAME type EOS, thermodynamic loci were calculated
with $\{\rho,e\}$ or $\{\rho,T\}$ primitive states, for comparison
(Fig.~\ref{fig:AlcmpT}).
For a monotonic EOS, the results were indistinguishable within differences from
forward or reverse interpolation of the tabular relations.
When the EOS, or the effective surface using a given order of interpolating
function, was non-monotonic, the results varied greatly because of
non-uniqueness when eliminating $T$ for the $\{\rho,e\}$ primitive state.

\subsubsection{Temperature model for mechanical equations of state}
Mechanical EOS are often available as empirical, algebraic relations 
$p(\rho,e)$, derived from shock data.
Temperature can be calculated without altering the mechanical EOS by adding
a relation $T(\rho,e)$.
While this relation could take any form in principle, one can also follow
the logic of the Gr\"uneisen EOS, in which the pressure is defined
in terms of its deviation $\Delta p(\rho,e-e_r)$ from
a reference curve $\{p_r,e_r\}(\rho)$.
Thus temperatures can be calculated by reference to a compression curve along
which the temperature and specific internal energy are known,
$\{T_r,e_r\}(\rho)$,
and a specific heat capacity defined as a function of density $c_v(\rho)$.
In the calculations, this augmented EOS was represented 
as a `mechanical-thermal' form comprising any $p(\rho,e)$ EOS plus the
reference curves -- an example of software inheritance and polymorphism.

One natural reference curve for temperature is the cold curve, $T_r=0$\,K.
The cold curve can be estimated from the principal isentrope
$\left.e(\rho)\right|_{s_0}$
using the estimated density variation of the Gr\"uneisen parameter:
\begin{equation}
e_r(\rho) = \left.e(\rho)\right|_{s_0} - T_0 c_p e^{a(1-\rho_0/\rho)}
   \left(\frac\rho{\rho_0}\right)^{\gamma_0-a}
\end{equation}
\cite{Steinberg96}.
In this work, the principal isentrope was calculated in tabular form 
from the mechanical EOS, using the ramp compression algorithm described above.

Empirical EOS are calibrated using experimental data.
Shock and adiabatic compression measurements on strong materials
inevitably include elastic-plastic contributions as well as the 
scalar EOS itself.
If the elastic-plastic contributions are not taken into account
self-consistently, the EOS may implicitly include contributions from
the strength.
A unique scalar EOS can be constructed to reproduce the normal stress
as a function of compression for any unique loading path:
shock or adiabat, for a constant or smoothly-varying strain rate.
Such an EOS would not generally predict the response to other
loading histories.
The EOS and constitutive properties for the materials
considered here were constructed self-consistently from shock data
-- this does not mean the models are accurate for other loading paths,
as neither the EOS nor the strength model includes all the physical terms
that real materials exhibit.
This does not in any case matter for the purposes of demonstrating the
properties of the numerical schemes.

This mechanical-thermal procedure was applied to Al using
a Gr\"uneisen EOS fitted to the same shock data used to calculate the 
$\{p,e\}(\rho,T)$ EOS discussed above \cite{Steinberg96}.
Temperatures were in good agreement (Fig.~\ref{fig:AlcmpT}).
The mechanical-thermal calculations required a similar computational effort to
the tabular $\{p,e\}(\rho,T)$ EOS with a $\{\rho,T\}$ primitive states
(and were thus much more efficient than the tabular EOS with  $\{\rho,e\}$
states), and described the EOS far more compactly.

\subsection{Strength}
For dynamic compressions to $o(10\,\mbox{GPa})$ and above,
on microsecond time scales,
the flow stress of solids is often treated as a correction or 
small perturbation to the scalar EOS.
However, the flow stress has been observed to be much higher on nanosecond
time scales \cite{Swift_pop_05},
and interactions between elastic and plastic waves may have a significant
effect on the compression and wave propagation.
The Rankine-Hugoniot equations should be solved self-consistently with
strength included.

\subsubsection{Preferred representation of isotropic strength}
There is an inconsistency in the standard continuum dynamics treatment
of scalar (pressure) and tensor (stress) response.
The scalar EOS expresses the pressure $p(\rho,e)$ as the
dependent quantity, which is the most convenient form for use in the
continuum equations.
Standard practice is to use sub-Hookean elasticity (hypoelastic form) 
\cite{Benson92} (Table~\ref{tab:stateexamples}), in which the state parameters
include the stress deviator $\sigma$, evolved by integration
\begin{equation}
\dot\sigma = G(s)\dot\epsilon
\end{equation}
where $G$ is the shear modulus and $\dot\epsilon$ the strain rate deviator.
Thus the isotropic and deviatoric contributions to stress are not
treated in an equivalent way: the pressure is calculated from a local state
involving a strain-like parameter (mass density), whereas the stress
deviator evolves with the time-derivative of strain.
This inconsistency causes problems along complicated loading paths
because $G$ varies strongly with compression:
if a material is subjected to a shear strain $\epsilon$, 
then isotropic compression (increasing the shear modulus from $G$ to $G'$, 
leaving $\epsilon$ unchanged),
then shear unloading to isotropic stress, the true unloading strain
is $-\epsilon$, whereas the hypoelastic calculation would require
a strain of $-\epsilon G/G'$.
Using Be and the Steinberg-Guinan strength model as an example of the
difference between hypoelastic and hyperelastic calculations, consider
an initial strain to a flow stress of 0.3\,GPa followed by
isothermal, isotropic compression to 100\,GPa,.
the strain to unload to a state of
isotropic stress is 0.20\%\ (hyperelastic) and 0.09\%\ (hypoelastic).
The discrepancy arises because the hypoelastic model does not increase
the deviatoric stress under compression at constant deviatoric strain.

The stress can be considered as a direct response of the
material to the instantaneous state of elastic strain: $\sigma(\epsilon,T)$.
This relation can be predicted directly with electronic structure calculations
of the stress tensor in a solid for a given compression and elastic strain
state \cite{Poirier99},
and is a direct generalization of the scalar equation of state.
A more consistent representation of the state parameters is to use
the strain deviator $\epsilon$ rather than $\sigma$, and to calculate
$\sigma$ from scratch when required using
\begin{equation}
\sigma = G(s)\epsilon
\end{equation}
-- a hyperelastic formulation.
The state parameters are then $\{\rho,e,\epsilon,\tilde\epsilon_p\}$.

The different formulations give different answers when deviatoric strain
is accumulated at different compressions, in which case 
the hyperelastic formulation is correct.
If the shear modulus varies with strain deviator -- {\it i.e.}, for nonlinear
elasticity -- then the definition of $G(\epsilon)$ must be adjusted
to give the same stress for a given strain.

Many isotropic strength models use scalar measures of the strain and stress
to parameterize work hardening and to apply a yield model of flow stress:
\begin{equation}
\tilde\epsilon=\sqrt{f_\epsilon||\epsilon^2||}, \quad
\tilde\sigma=\sqrt{f_\sigma||\sigma^2||}.
\end{equation}
Inconsistent conventions for equivalent scalar measures have been used by
different workers.
In the present work, the common shock physics convention was used that
the flow stress component of $\tau_n$ is $\frac 23Y$ 
where $Y$ is the flow stress.
For consistency with published speeds and amplitudes for elastic waves,
$f_\epsilon=f_\sigma=\frac 32$, in contrast to other values previously
used for lower-rate deformation \cite{Hill}.
In principle, the values of $f_\epsilon$ and $f_\sigma$ do not matter
as long as the strength parameters were calibrated using the same values
then used in any simulations.

\subsubsection{Beryllium}
The flow stress measured from laser-driven shock experiments on
Be crystals a few tens of micrometers thick is, at around 5-9\,GPa 
\cite{Swift_pop_05},
much greater than the 0.3-1.3\,GPa measured on microsecond time scales.
A time-dependent crystal plasticity model for Be is being developed,
and the behavior under dynamic loading depends on the detailed time dependence
of plasticity.
Calculations were performed with the Steinberg-Guinan strength model
developed for microsecond scale data \cite{Steinberg96},
and, for the purposes of rough comparison, with
elastic-perfectly plastic response with a flow stress of 10\,GPa.
The elastic-perfectly plastic model neglected pressure- and work- hardening.

\begin{figure}
\begin{center}\includegraphics[scale=0.80]{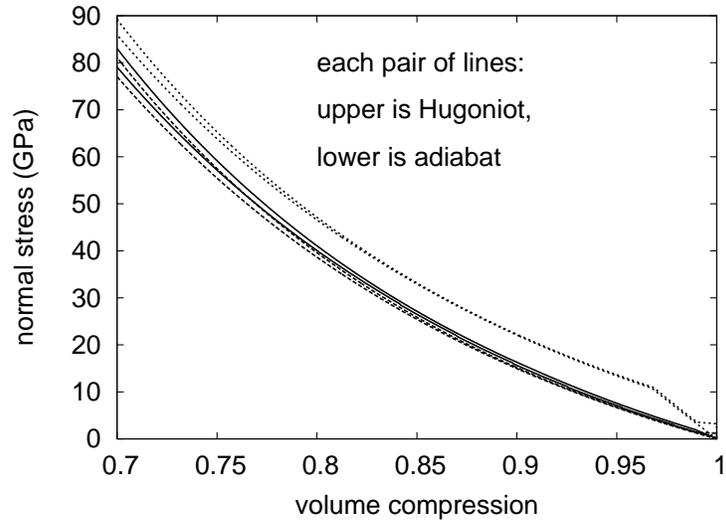}\end{center}
\caption{Principal adiabat and shock Hugoniot
   for Be in normal stress-compression space,
   neglecting strength (dashed), for Steinberg-Guinan strength (solid),
   and for elastic-perfectly plastic with $Y=10$\,GPa (dotted).}
\label{fig:Becmpvp}
\end{figure}

Calculations were made of the principal adiabat and shock Hugoniot,
and of a release adiabat from a state on the principal Hugoniot.
Calculations were made with and without strength.
Considering the state trajectories in stress-volume space,
it is interesting to note that heating from plastic flow may push 
the adiabat above the Hugoniot, because of the greater heating obtained by
integrating along the adiabat compared with jumping from the initial to the
final state on the Hugoniot (Fig.~\ref{fig:Becmpvp}).
Even with an elastic-perfectly plastic strength model, the with-strength
curves do not lie exactly $\frac 23Y$ above the strengthless curves,
because heating from plastic flow contributes an increasing amount of
internal energy to the EOS as compression increases.

\begin{figure}
\begin{center}\includegraphics[scale=0.80]{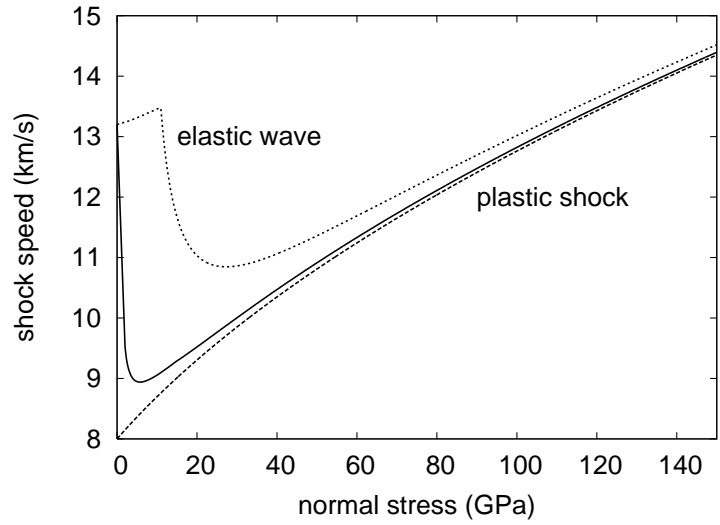}\end{center}
\caption{Principal adiabat and shock Hugoniot
   for Be in shock speed-normal stress space,
   neglecting strength (dashed), for Steinberg-Guinan strength (solid),
   and for elastic-perfectly plastic with $Y=10$\,GPa (dotted).}
\label{fig:Becmppus}
\end{figure}

An important characteristic for the seeding of instabilities by
microstructural variations in shock response is the shock stress at which 
an elastic wave does not run ahead of the shock.
In Be with the high flow stress of nanosecond response,
the relation between shock and particle speeds is significantly different
from the relation for low flow stress (Fig.~\ref{fig:Becmppus}).
For low flow stress, the elastic wave travels at 13.2\,km/s.
A plastic shock travels faster than this for pressures greater than 110\,GPa,
independent of the constitutive model.
The speed of a plastic shock following the initial elastic wave is similar to 
the low strength case, because the material is already at its flow stress,
but the speed of a single plastic shock is appreciably higher.

\begin{figure}
\begin{center}\includegraphics[scale=0.80]{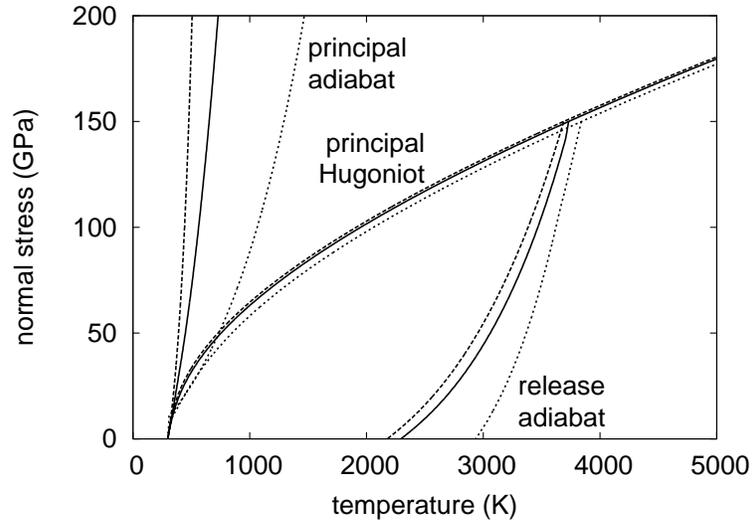}\end{center}
\caption{Principal adiabat, shock Hugoniot, and release adiabat
   for Be in normal stress-temperature space,
   neglecting strength (dashed), for Steinberg-Guinan strength (solid),
   and for elastic-perfectly plastic with $Y=10$\,GPa (dotted).}
\label{fig:Becmptp}
\end{figure}

For compression to a given normal stress, the temperature is significantly
higher with plastic flow included.
The additional heating is particularly striking on the principal adiabat:
the temperature departs significantly from the principal isentrope.
Thus ramp-wave compression of strong materials may lead to significant
levels of heating, contrary to common assumptions of small temperature
increases \cite{Hall00}.
Plastic flow is largely irreversible, so heating occurs on unloading
as well as loading.
Thus, on adiabatic release from a shock-compressed state, additional
heating occurs compared with the no-strength case.
These levels of heating are important as shock or release melting may
occur at a significantly lower shock pressure than would be expected
ignoring the effect of strength.
(Fig.~\ref{fig:Becmptp}.)

\subsection{Phase changes}
An important property of condensed matter is phase changes,
including solid-solid polymorphism and solid-liquid.
An equilibrium phase diagram can be represented as a single overall EOS
surface as before.
Multiple, competing phases with kinetics for each phase transformation
can be represented conveniently using
the structure described above for general material properties, 
for example by describing the local state as a set of volume fractions
$f_i$ of each possible simple-EOS phase, with transition rates and
equilibration among them.
This model is described in more detail elsewhere \cite{rfsim}.
However, it is interesting to investigate the robustness of the 
numerical scheme for calculating shock Hugoniots when the EOS has
the discontinuities in value and gradient associated with phase changes.

\begin{figure}
\begin{center}\includegraphics[scale=0.80]{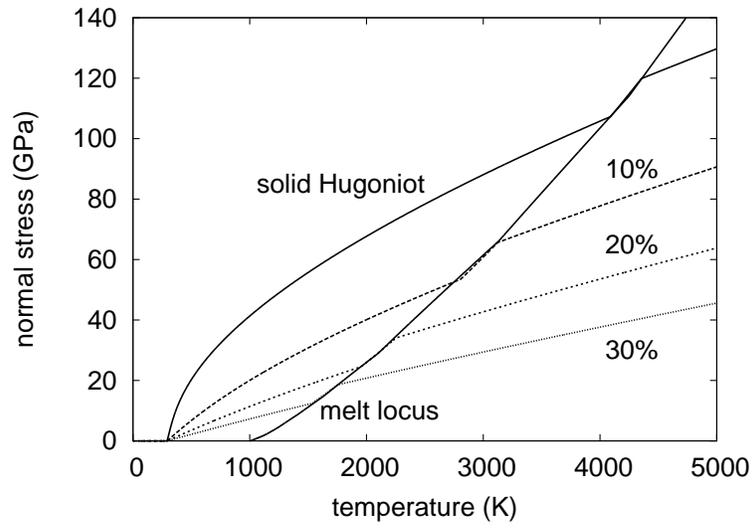}\end{center}
\caption{Demonstration of shock Hugoniot solution across a phase boundary:
   shock-melting of Al, for different initial porosities.}
\label{fig:Almelt}
\end{figure}

The EOS of molten metal, and the solid-liquid phase transition, can be
represented to a reasonable approximation as an adjustment to the EOS of the
solid:
\begin{equation} 
p_{\mbox{two-phase}}(\rho,e) = p_{\mbox{solid}}(\rho,\tilde e)
\end{equation}
where
\begin{equation}
\tilde e = 
\left\{\begin{array}{ll}
e & \quad : \quad T(\rho,e) < T_m(\rho) \\
e-\Delta\tilde h_m & \quad : \quad \Delta\tilde h_m\equiv c_v(\rho,e)\left[T(\rho,e)-T_m(\rho)\right] < \Delta h_m \\
e-\Delta h_m & \quad : \quad \mbox{otherwise} \\
\end{array}
\right.
\end{equation}
and $\Delta h_m$ is the specific latent heat of fusion.
Taking the EOS and a modified Lindemann melting curve for Al
\cite{Steinberg96}, and using $\Delta h_m=0.397$\,MJ/kg,
the shock Hugoniot algorithm was found to operate stably across the phase
transition (Fig.~\ref{fig:Almelt}).

\section{Composite loading paths}
Given methods to calculate shock and adiabatic loading paths from arbitrary
initial states,
a considerable variety of experimental scenarios can be treated
from the interaction of loading or unloading waves with interfaces
between different materials,
in planar geometry for uniaxial compression.
The key physical constraint is that, if two dissimilar materials are to
remain in contact after an interaction such as an impact or the passage of
a shock, the normal stress $\tau_n$ and particle speed $u_p$ in
both materials must be equal on either side of the interface.
The change in particle speed and stress normal to the waves were 
calculated above for compression waves running in the direction of
increasing spatial ordinate (left to right).
Across an interface, the sense is reversed for the material at the left.
Thus a projectile impacting a stationary target to the right is
decelerated from its initial speed by the shock induced by impact.

\begin{figure}
\begin{center}\includegraphics[scale=0.60]{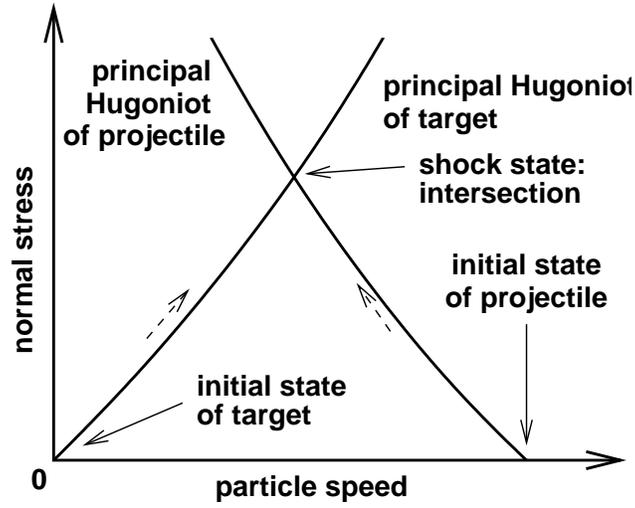}\end{center}
\caption{Wave interactions for the impact of a flat projectile
   moving from left to right with a stationary target.
   Dashed arrows are a guide to the sequence of states.
   For a projectile moving from right to left, the construction is the
   mirror image reflected in the normal stress axis.}
\label{fig:impact}
\end{figure}

\begin{figure}
\begin{center}\includegraphics[scale=0.60]{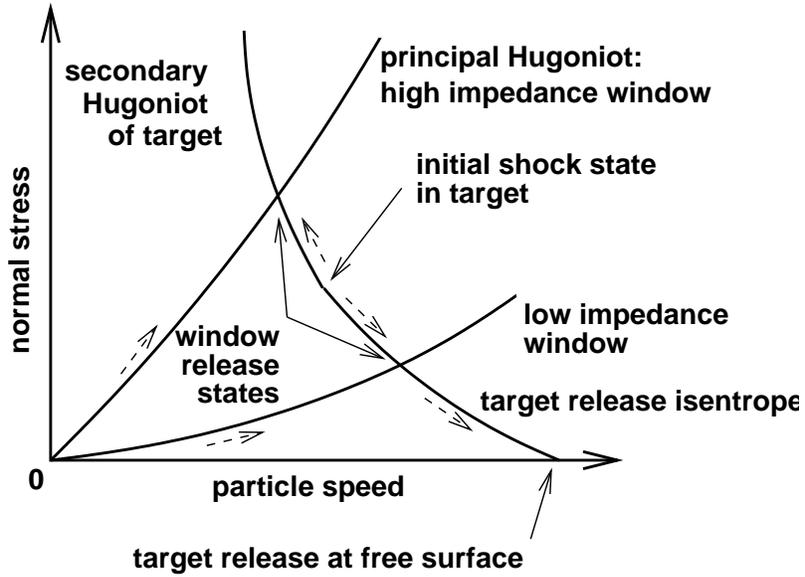}\end{center}
\caption{Wave interactions for the release of a shocked state 
   (shock moving from left to right)
   into a stationary `window' material to its right.
   The release state depends whether the window has a higher or lower
   shock impedance than the shocked material.
   Dashed arrows are a guide to the sequence of states.
   For a shock moving from right to left, the construction is the
   mirror image reflected in the normal stress axis.}
\label{fig:shockrel}
\end{figure}

\begin{figure}
\begin{center}\includegraphics[scale=0.60]{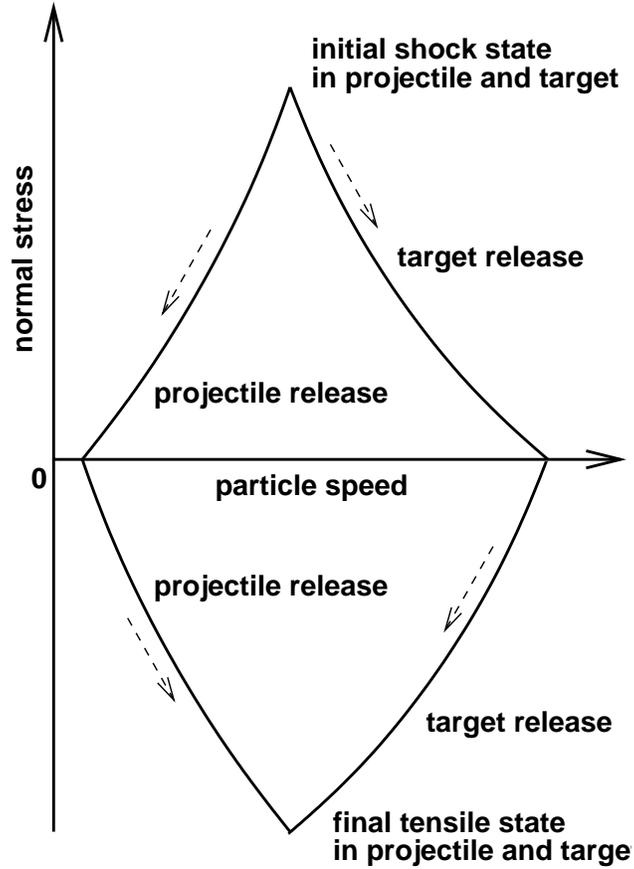}\end{center}
\caption{Wave interactions for the release of a shocked state 
   by tension induced as materials try to separate in opposite directions
   when joined by a bonded interface.
   Material damage, spall, and separation are neglected: the construction
   shows the maximum tensile stress possible.
   For general material properties, {\it e.g.} if plastic flow is included,
   the state of maximum tensile stress is not just the negative of the
   initial shock state.
   Dashed arrows are a guide to the sequence of states.
   The graph shows the initial state after an impact by a projectile moving
   from right to left;
   for a shock moving from right to left, the construction is the
   mirror image reflected in the normal stress axis.}
\label{fig:doublerel}
\end{figure}

\begin{figure}
\begin{center}\includegraphics[scale=0.60]{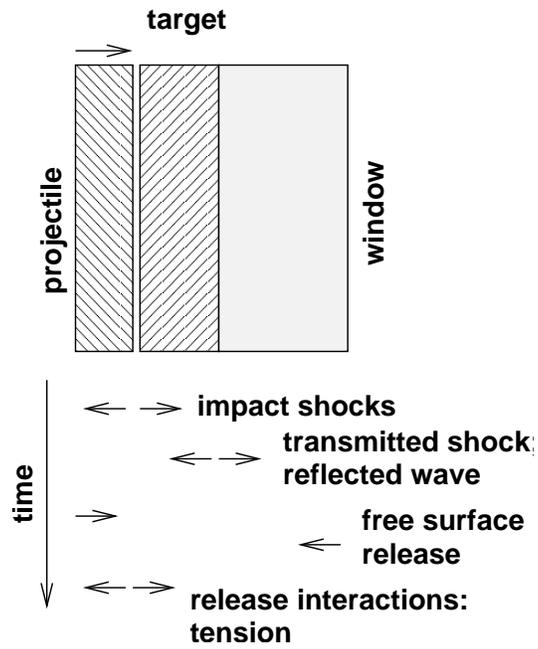}\end{center}
\caption{Schematic of uniaxial wave interactions induced by the impact
   of a flat projectile with a composite target.}
\label{fig:composite}
\end{figure}

The general problem at an interface can be analyzed by considering the
states at the instant of first contact -- on impact, or when a shock 
traveling through a sandwich of materials first reaches the interface.
The initial states are $\{u_l,s_l;u_r,s_r\}$.
The final states are $\{u_j,s_l';u_j,r_r'\}$ where $u_j$ is the
joint particle speed, $\tau_n(s_l')=\tau_n(s_r')$,
and $s_i'$ is connected to $s_i$ by either a shock or an adiabat,
starting at the appropriate initial velocity and stress,
and with orientation given by the side of the system each material
occurs on.
Each type of wave is considered in turn, looking for an intersection
in the $u_p-\tau_n$ plane.
Examples of these wave interactions are
the impact of a projectile with a stationary target (Fig.~\ref{fig:impact}),
release of a shock state at a free surface or a material
({\it e.g.} a window) of lower
shock impedance (hence reflecting a release wave into the shocked
material -- Fig.~\ref{fig:shockrel}), 
reshocking at a surface with a material of higher shock impedance
(Fig.~\ref{fig:shockrel}),
or tension induced as materials try to separate in opposite directions
when joined by a bonded interface (Fig.~\ref{fig:doublerel}).
Each of these scenarios may occur in turn following the impact of
a projectile with a target: if the target is layered then a shock is
transmitted across each interface with a release or a reshock reflected back,
depending on the materials;
release ultimately occurs at the rear of the projectile and the far end of
the target, and the oppositely-moving release waves subject the projectile
and target to tensile stresses when they interact
(Fig.~\ref{fig:composite}).

\begin{figure}
\begin{center}\includegraphics[scale=0.80]{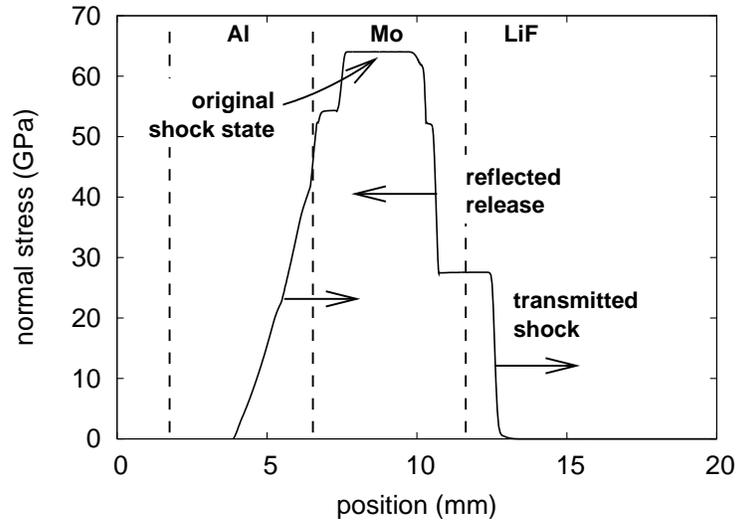}\end{center}
\caption{Hydrocode simulation of Al projectile at 3.6\,km/s
   impacting a Mo target with a LiF release window,
   1.1\,$\mu$s after impact.
   Structures on the waves are elastic precursors.}
\label{fig:impactrelcmp}
\end{figure}

As an illustration of combining shock and ramp loading calculations,
consider the problem of an Al projectile, initially traveling at 3.6\,km/s,
impacting a stationary, composite target 
comprising a Mo sample and a LiF release window
\cite{Swift_Mo_07,Seifter08}.
The shock and release states were calculated 
using published material properties \cite{Steinberg96}.
The initial shock state was calculated to have a normal stress of
63.9\,GPa.
On reaching the LiF, the shock was calculated to transmit at 27.1\,GPa,
reflecting as a release in the Mo.
These stresses match the continuum dynamics simulation to within 0.1\,GPa
in the Mo and 0.3\,GPa in the LiF, using the same material properties
(Fig.~\ref{fig:impactrelcmp}).
The associated wave and particle speeds match to a similar accuracy;
wave speeds are much more difficult to extract from the continuum dynamics
simulation.

An extension of this analysis can be used to calculate the
interaction of oblique shocks with an interface \cite{Loomis07}.

\section{Conclusions}
A general formulation was developed to represent material models 
for applications in dynamic loading, suitable for software implementation
in object-oriented programming languages.
Numerical methods were devised to calculate the response of matter
represented by the general material models to shock and ramp compression,
and ramp decompression,
by direct evaluation of the thermodynamic pathways for these compressions
rather than spatially-resolved simulations.
This approach is a generalization of earlier work on solutions for materials
represented by a scalar equation of state.
The numerical methods were found to be flexible and robust: capable of
application to materials with very different properties.
The numerical solutions matched analytic results to a high accuracy.

Care was needed with the interpretation of some types of physical response,
such as plastic flow, when applied to deformation at high strain rates.
The underlying time-dependence of processes occurring during deformation
should be taken into account.
The actual history of loading and heating experienced by material 
during the passage of a shock may influence the final state -- this history
is not captured in the continuum approximation to material dynamics,
where shocks are treated as discontinuities.
Thus care is also needed in spatially resolved simulations when shocks
are modeled using artificial viscosity to smear them unphysically over
a finite thickness.

Calculations were shown to demonstrate the operation of the algorithms
for shock and ramp compression with material models representative of
complex solids including strength and phase transformations.

The basic ramp and shock solution methods were coupled to solve for
composite deformation paths, such as shock-induced impacts, and shock
interactions with a planar interface between different materials.
Such calculations capture much of the physics of typical material dynamics
experiments, without requiring spatially-resolving simulations.
The results of direct solution of the relevant shock and ramp loading 
conditions were compared with hydrocode simulations, showing complete
consistency.

\section*{Acknowledgments}
Ian Gray introduced the author to the concept of 
multi-model material properties software.
Lee Markland developed a prototype Hugoniot-calculating computer
program for equations of state while working for the author as an undergraduate
summer student.

Evolutionary work on material properties libraries was supported
by the U.K. Atomic Weapons Establishment, Fluid Gravity Engineering Ltd,
and Wessex Scientific and Technical Services Ltd.
Refinements to the technique and applications to the problems described
were undertaken at Los Alamos National Laboratory (LANL) and
Lawrence Livermore National Laboratory (LLNL).

The work was performed partially in support of, and funded by,
the National Nuclear Security Agency's
Inertial Confinement Fusion program at LANL (managed by Steven Batha), and
LLNL's Laboratory-Directed Research and Development project 06-SI-004
(Principal Investigator: Hector Lorenzana).
The work was performed under the auspices of
the U.S. Department of Energy under contracts W-7405-ENG-36,
DE-AC52-06NA25396, and DE-AC52-07NA27344.



\clearpage
\begin{table}
\caption{Interface to material models required for explicit forward-time
   continuum dynamics simulations.}
\label{tab:interface}
\begin{center}
\begin{tabular}{ll}\hline
{\bf purpose} & {\bf interface calls} \\ 
\hline
program set-up & read/write material data \\
\hline
continuum dynamics equations & stress(state) \\
time step control & sound speed(state) \\
\hline
evolution of state (deformation) & $d(\mbox{state})/dt$(state,$\grad\,\vec u$) \\
evolution of state (heating) & $d(\mbox{state})/dt$(state,$\dot e$) \\
internal evolution of state & $d(\mbox{state})/dt$ \\
\hline
manipulation of states & create and delete \\
 & add states \\
 & multiply state by a scalar \\
 & check for self-consistency \\
\hline\end{tabular}
\end{center}

Parentheses in the interface calls denote functions,
{\it e.g.} ``stress(state)'' for
``stress as a function of the instantaneous, local state.''
The evolution functions are shown in the operator-split structure that is
most robust for explicit, forward-time numerical solutions and can also be
used for calculations of the shock Hugoniot and ramp compression.
Checks for self-consistency include that mass density is positive, volume
or mass fractions of components of a mixture add up to one, etc.
\end{table}

\clearpage
\begin{table}
\caption{Examples of types of material model, distinguished by different
structures in the state vector.}
\label{tab:stateexamples}
\begin{center}
\begin{tabular}{lll}\hline
{\bf model} & {\bf state vector} & {\bf effect of mechanical strain} \\
 & $s$ & $\dot s_m(s,\grad u)$ \\ \hline
mechanical equation of state & $\rho,e$ & $-\rho\div\vec u,-p\div\vec u/\rho$ \\
thermal equation of state & $\rho,T$ & $-\rho\div\vec u,-p\div\vec u/\rho c_v$ \\
heterogeneous mixture & $\{\rho,e,f_v\}_i$ & $\{-\rho\div\vec u,-p\div\vec u/\rho,0\}_i$ \\
homogeneous mixture & $\rho,T,\{f_m\}_i$ & $\{-\rho\div\vec u,-p\div\vec u/\rho c_v,0_i$ \\
traditional deviatoric strength & $\rho,e,\sigma,\tilde\epsilon_p$ & $-\rho\div\vec u,\frac{-p\div\vec u+f_p||\sigma\dot\epsilon_p||}\rho,G\dot\epsilon_e,$ \\
 &  & $\sqrt{f_\epsilon||\dot\epsilon_p^2||}$ \\
\hline\end{tabular}
\end{center}

The symbols are $\rho$: mass density; $e$: specific internal energy, 
$T$: temperature, 
$f_v$: volume fraction, $f_m$: mass fraction,
$\sigma$: stress deviator,
$f_p$: fraction of plastic work converted to heat,
$\grad u_p$: plastic part of velocity gradient,
$G$: shear modulus,
$\dot\epsilon_{e,p}$: elastic and plastic parts of strain rate deviator,
$\tilde\epsilon_p$: scalar equivalent plastic strain,
$f_\epsilon$: factor in effective strain magnitude.
Reacting solid explosives can be represented as heterogeneous mixtures,
one component being the reacted products;
reaction, a process of internal evolution, transfers material from 
unreacted to reacted components.
Gas-phase reaction can be represented as a homogeneous mixture,
reactions transferring mass between components representing different
types of molecule.
Symmetric tensors such as the stress deviator are represented more compactly
by their 6 unique upper triangular components, {\it e.g.} using Voigt notation.
\end{table}

\clearpage
\begin{table}
\caption{Outline hierarchy of material models, illustrating the use of
   polymorphism (in the object-oriented programming sense).}
\label{tab:polymorphism}
\begin{center}
\begin{tabular}{lp{2.5in}}\hline
{\bf material (or state) type} & {\bf model type} \\
\hline
mechanical equation of state &
   polytropic, Gr\"uneisen,
   energy-based Jones-Wilkins-Lee, $(\rho,T)$ table, etc \\
thermal equation of state & temperature-based Jones-Wilkins-Lee,
 quasiharmonic, $(\rho,T)$ table, etc \\
reactive equation of state & modified polytropic, reactive Jones-Wilkins-Lee \\
spall & Cochran-Banner \\
deviatoric stress &  elastic-plastic, Steinberg-Guinan, Steinberg-Lund,
    Preston-Tonks-Wallace, etc \\
homogeneous mixture & mixing and reaction models \\
heterogeneous mixture & equilibration and reaction models \\
\hline\end{tabular}
\end{center}

Continuum dynamics programs can refer to material properties as an
abstract `material type' with an abstract material state.
The actual type of a material ({\it e.g.} mechanical equation of state),
the specific model type ({\it e.g.} polytropic), 
and the state of material of that
type are all handled transparently by the object-oriented software structure.

The reactive equation of state has an additional state parameter $\lambda$,
and the software operations are defined by extending those of the
mechanical equation of state.
Spalling materials can be represented by a solid state plus a void fraction
$f_v$, with operations defined by extending those of the solid material.
Homogeneous mixtures are defined as a set of thermal equations of state,
and the state is the set of states and mass fractions for each.
Heterogeneous mixtures are defined as a set of `pure' material properties
of any type, and the state is the set of states for each component plus
its volume fraction.
\end{table}

\clearpage

\clearpage

\clearpage

\clearpage

\clearpage

\clearpage

\clearpage

\clearpage

\clearpage

\clearpage

\clearpage

\clearpage
\section*{List of figures}
\begin{enumerate}

\item{Principal isentrope and shock Hugoniot for air (perfect gas):
   numerical calculations for general material models,
   compared with analytic solutions.}

\item{Shock Hugoniot for Al in pressure-temperature space,
   for different representations of the equation of state.}

\item{Principal adiabat and shock Hugoniot
   for Be in normal stress-compression space,
   neglecting strength (dashed), for Steinberg-Guinan strength (solid),
   and for elastic-perfectly plastic with $Y=10$\,GPa (dotted).}

\item{Principal adiabat and shock Hugoniot
   for Be in shock speed-normal stress space,
   neglecting strength (dashed), for Steinberg-Guinan strength (solid),
   and for elastic-perfectly plastic with $Y=10$\,GPa (dotted).}

\item{Principal adiabat, shock Hugoniot, and release adiabat
   for Be in normal stress-temperature space,
   neglecting strength (dashed), for Steinberg-Guinan strength (solid),
   and for elastic-perfectly plastic with $Y=10$\,GPa (dotted).}

\item{Demonstration of shock Hugoniot solution across a phase boundary:
   shock-melting of Al, for different initial porosities.}

\item{Wave interactions for the impact of a flat projectile
   moving from left to right with a stationary target.
   Dashed arrows are a guide to the sequence of states.
   For a projectile moving from right to left, the construction is the
   mirror image reflected in the normal stress axis.}

\item{Wave interactions for the release of a shocked state 
   (shock moving from left to right)
   into a stationary `window' material to its right.
   The release state depends whether the window has a higher or lower
   shock impedance than the shocked material.
   Dashed arrows are a guide to the sequence of states.
   For a shock moving from right to left, the construction is the
   mirror image reflected in the normal stress axis.}

\item{Wave interactions for the release of a shocked state 
   by tension induced as materials try to separate in opposite directions
   when joined by a bonded interface.
   Material damage, spall, and separation are neglected: the construction
   shows the maximum tensile stress possible.
   For general material properties, {\it e.g.} if plastic flow is included,
   the state of maximum tensile stress is not just the negative of the
   initial shock state.
   Dashed arrows are a guide to the sequence of states.
   The graph shows the initial state after an impact by a projectile moving
   from right to left;
   for a shock moving from right to left, the construction is the
   mirror image reflected in the normal stress axis.}

\item{Schematic of uniaxial wave interactions induced by the impact
   of a flat projectile with a composite target.}

\item{Hydrocode simulation of Al projectile at 3.6\,km/s
   impacting a Mo target with a LiF release window,
   1.1\,$\mu$s after impact.
   Structures on the waves are elastic precursors.}
\end{enumerate}

\end{document}